\documentclass{llncs}

\usepackage{amsmath}
\usepackage{amssymb}
\usepackage{amsfonts}

\usepackage{stmaryrd}

\usepackage{graphicx}

\newcommand{\eqdef}{:=}
\newcommand{\F}{\mathbf{F}}
\newcommand{\Z}{\mathbf{Z}}

\newcommand{\xx}{\mathbf{x}}

\newcommand{\yy}{\mathbf{y}}
\newcommand{\zero}{\mathbf{0}}
\newcommand{\uu}{\mathbf{u}}
\newcommand{\vv}{\mathbf{v}}

\date{}

\usepackage[bookmarks=false]{hyperref}
\hypersetup{
  pdfhighlight=/O,
  linkcolor=black, citecolor=black, filecolor=black, urlcolor=black,
  colorlinks=true,
  breaklinks=true,
  bookmarksopen=true,
  bookmarksopenlevel=0,
  pdfauthor={Frédéric Didier and Yann Laigle-Chapuy},
  pdftitle={Attacking the combination generator},
  pdfsubject={},
  pdfkeywords={cryptography, correlation attack, combination generator}
}

\title{Attacking the combination generator
\thanks{This work is partially funded by CELAR/DGA.}
}
\author{Fr\'ed\'eric Didier\inst{1} and Yann Laigle-Chapuy\inst{2}}
\institute{
EPFL, IC - IIF - ALGO, B\^atiment BC,\\
Station 14, CH - 1015 Lausanne\\
\email{frederic.didier@epfl.ch}\\
\and Projet SECRET, INRIA Rocquencourt, Domaine de Voluceau,\\
78153 Le Chesnay cedex\\
\email{yann.Laigle-Chapuy@inria.fr}
}

\begin{document}

\maketitle

\begin{abstract}
We present one of the most efficient attacks against the combination generator.
This attack is inherent to this system as its only assumption is that the filtering function has a good autocorrelation.
This is usually the case if the system is designed to be resistant to other kinds of attacks.
We use only classical tools, namely vectorial correlation, weight 4 multiples and Walsh transform.
\end{abstract}

\vspace{0.5cm}
\noindent
{\bf Keywords}:
Stream cipher, combination generator, Boolean functions, low weight multiples, Walsh transform.

\section{Introduction}\label{intro}

The combination generator is, together with the filter generator, one
of the simplest and most analyzed construction of stream ciphers. It
uses as an internal state many linear feedback shift registers
(LFSRs).
We will write $m$ for their total size in bits.
These registers are filtered using a
$n$-variable balanced Boolean function $f$ (from $\F_2^n$ into
$\F_2$) to produce the keystream $(z_t)_{t\ge 0}$. The inputs of this
function are taken from some bits in the LFSRs internal states. We will
write $\xx_t$ for the $n$-bit vector corresponding to the inputs of
$f$ at time $t$. Notice that we will always write such vector of bits
in bold. Our goal here is to find the key (that is the initial state
of all the LFSRs) knowing the keystream sequence $(z_t)_{t\ge0}$ and
all the components of the combination generator.


The classical way to attack such a system is to use a correlation attack \cite{Sieg85} or
one of its variants called fast correlation attacks \cite{MS88,CT00,JJ00}.
The idea is to exploit the existence of a statistical dependence between the keystream
and one of the constituent LFSR. For example, if we write $\xx=(\uu,\vv)$ where $\uu$
corresponds to the bits taken from the first LFSR, such a dependence exists if
for some linear function $\ell$,
$$ \Pr[f(\uu,\vv) = \ell(\uu) ] \ne 1/2 \ .$$
One can then perform an exhaustive search on the initial state of the first LFSR (or any
other targeted LFSR) and try to detect a bias. In order to ensure that there is no such
bias (or a really low one) $f$ is usually chosen with a high non-linearity and a
low autocorrelation.

In the fast correlation attacks, one sees the search of the initial
state as a decoding problem of the linear code generated by the
target LFSR and the linear function $\ell$. Then, instead of an
exhaustive search, it is possible to avoid examining all possible
initializations of the target LFSR by using some efficient
error-correcting techniques. Usually the attack becomes faster but
requires a larger amount of keystream.


This paper presents a new attack on the combination generator. At
first sight, it may not seems that new because we use only classical
tools, namely low weight multiples, vectorial correlation and Walsh
transform. But in our knowledge, they are used in an original way and
we add some interesting insights. Moreover, this attack is not based
on any particular weakness of the filtering function like for
correlation attacks. Actually, we can even attack a system where we
do not know the filtering function used! To simplify our analysis we
only assume that $f$ has a good autocorrelation. However choosing a
filtering function without this property will open the door to the
correlation attacks described above.

The paper is organized as follows. We begin by explaining the attack
principle in the next section. Then, in Section 3, we detail the
algorithm and the complexity of the different steps involved. We give
in Section 4 the actual time complexity of our implementation on an
example combination generator. We finally conclude in the last
section.

\section{Attack principle}\label{principle}

We describe here an attack on the combination generator based on the theorem given at the end of this section.
What we actually show is how to find the initial state of some of the LFSR composing the system.
This is certainly enough to say that the system is insecure and recovering the rest of the initial state require usually less work.
The best method for this task is more dependent on the actual system we are working on and we will give in Section \ref{example} an example of how we can actually do it.

So, suppose we are given a combination generator and we are able to
observe a large amount of keystream bits, the $z_t$'s. We denote by $\xx_t$
the input of the filtering function for this observed system at time
$t$, that is we have $z_t=f(\xx_t)$. We split the LFSRs involved into
two groups and do an exhaustive search on the initial states of the
first LFSRs. For a given value $I$ of this partial initial state, we
write $\yy_t$ for the hypothetical first part of the filtering
function input at time $t$. That is, the part we can compute given
$I$. We also need one or more common weight 4 multiples $1+X^{t_1} +
X^{t_2} + X^{t_3}$ of the feedback polynomials of the LFSRs in the
second group. We then compute the ratio $P_I$ of times $t$ such that
$$ z_t + z_{t+t_1} + z_{t+t_2} + z_{t+t_3} = 0$$
amongst all the $t$'s such that
$$ \yy_t + \yy_{t+t_1} + \yy_{t+t_2} + \yy_{t+t_3} = \zero \ .$$
Remark that the number of such time instants at our disposal is directly linked to the amount of keystream we know.
Now, how can we distinguish the real partial initial state from the others?
\begin{itemize}
\item
If $I$ is the actual first part of the initial state, then we know we
look only at times $t$ such that $\xx_t+\xx_{t+t_1} + \xx_{t+t_2} +
\xx_{t+t_3} = \zero$. This is because we are working with a multiple
of the other LFSRs feedback polynomials. Then, it is reasonable to
assume that the ratio $P_I$ is an estimate of the probability
$\Pr(f(\uu_1)+f(\uu_2)+f(\uu_3)+f(\uu_4) = 0 \mid
\sum_i\uu_i = \zero)$. Here the $\uu_i$ are vectors in $\F_2^n$
uniformly distributed amongst the one having the required property.

\item
If $I$ is not the actual first part of the initial state, even in the worst case where only the initial state of one LFSR was not guessed correctly, it is reasonable to assume that the sum $\xx_t+\xx_{t+t_1} + \xx_{t+t_2} + \xx_{t+t_3}$ is different from $\zero$ roughly half of the time.
For theses points, we then have an estimate for one of the probabilities $\Pr(f(\uu_1)+f(\uu_2)+f(\uu_3)+f(\uu_4) = 0 \mid \sum_i \uu_i = \uu)$ where $\uu \ne \zero$.
\end{itemize}
Hopefully for us, the two cases are distinguishable thanks to the
following result taken almost directly from \cite{Did07} and already
present in the work of Sabine Leveiller \cite{Lev04}.

\begin{theorem}
\label{thm}
Let $f$ be a $n$-variable balanced Boolean function, and let $\uu_1$, $\uu_2$, $\uu_3$, $\uu_4$ be 4 uniformly distributed $n$ bits vector such that $\uu_1 + \uu_2 + \uu_3 + \uu_4 = \uu$.
Let $P_\uu \eqdef \Pr(f(\uu_1)+f(\uu_2)+f(\uu_3)+f(\uu_4) = 0)$, then we have
\begin{equation*}
P_\zero \ge \frac{1}{2} + \frac{1}{2^{n+1}}
\end{equation*}
and
\begin{equation*}
\min_{\uu \ne \zero} \left(P_0 - P_\uu\right) \ge \frac{1}{2^{n+1}} \left(1-\frac{\Delta_f}{2^n}\right)^2
\end{equation*}
where $\Delta_f$ is the maximum of the autocorrelation coefficients of $f$ and is usually small compared to $2^n$.
\end{theorem}
\begin{proof}
see Appendix.
\end{proof}

We will assume in the rest of this paper that the filtering function has a good autocorrelation property which result in a difference between these probabilities of $\frac{1}{2^{n+1}}$.
This is a reasonable hypothesis because it is the case for the function usually used in this settings \cite{GK03}.
In particular, it is a needed property to resist against correlation attacks.

\section{Detailed analysis}\label{analysis}

We describe here our attack in details and hence need some more notations.
Since we split the LFSRs in two groups, we will write $m=m_1+m_2$ where $m_1$ is the total size in bits of the LFSRs in the first group.
That is the one we do an exhaustive search on.
Similarly, we write $n=n_1+n_2$ for the inputs bit of the filtering function $f$.
Remark that in practice, $n$ is kept small compared to $m$ for efficiency issues.
Our result is summarized in this theorem:

\begin{theorem}
Our attack recovers $m_1$ bits of the initial states in complexity $O(m_1 2^{n_1} 2^{m_1})$.
It requires $O(2^\frac{m_2}{3} + m_1 2^{2n+n_1+1})$ consecutive bits of keystream and a
memory of $O(2^{m_1})$.
If the memory requirement is too big, some tradeoff exists until a complexity of $O(m_1 2^{2n+n_1} 2^{m_1})$ and a memory of $O(m_1 2^{2n+n_1})$.
This attack also require a precomputation phase of complexity and memory in $O(2^{\frac{m_2}{3}})$.
\end{theorem}

\subsection{Computing weight 4 multiples}

We need to compute one or a few weight 4 common multiples of the second group of LFSRs.
We show here that this precomputation phase can be dealt with a complexity and memory around
$O(2^{\frac{m_2}{3}})$.

What we need is only a few multiple of degree as small as possible.
If we look at LFSRs of total size $m_2$, it is well known that the
expected number of weight 4 multiples of degree $D$ is heuristically
approximated by $\frac{1}{2^{m_2}}\frac{D^3}{6}$ considering that for $D$
large enough the values of the polynomials of weight $4$ and degree
at most $D$ are uniformly distributed.

There are many algorithms to compute low weight multiples whose
complexity depends on the parameters $D$ which in our case is around
$2^{\frac{m_2}{3}}$. One can use the algorithm in \cite{CJM02} but
for weight 4 multiples, the most efficient is the one of
\cite{DLC07}. It's especially adapted here, as we need to find a
simultanous multiple of many polynomials.

Its complexity is in $O(D)$ times the complexity to
compute discrete logarithms in the multiplicative group
$\F_{2^{l_1}}^* \times \dots \times \F_{2^{l_k}}^*$ where the $l_i$
are the respective length of each LFSR. This task is particularly
easy since using Pohlig-Hellman algorithm it can be splited into $k$
discrete logarithms computation in each of the finite field
multiplicative groups involved.


\subsection{Amount of keystream needed}

We prove here the following lemma

\begin{lemma}\label{lemma1}
We need to consider around $N\eqdef m_1 2^{2n+n1+1}$ degree 4 equations to identify the correct partial initial state.
This translates to $O(2^\frac{m_2}{3} + m_1 2^{2n+n_1+1})$ consecutive keystream bits needed.
\end{lemma}

Given the results in Section \ref{principle}, mainly Theorem
\ref{thm}, we need to distinguish in the worst case between two
binomial distributions, one of parameter
$\frac{1}{2}+\frac{1}{2^{n+1}}$ and one of parameter at most
$\frac{1}{2} + \frac{1}{2^{n+2}}$. However, for most of the wrong
partial initial states, we will just observe a law of parameter
$\frac{1}{2}$, so the bias we need to detect is in practice close to
$\frac{1}{2^{n+1}}$.

A classical results from statistics tell us that using $S$ samples,
we have an error probability of roughly $2^{-\frac{S}{2^{2n+1}}}$. We
thus need at least $2^{2n+1}$ samples to be able to
distinguish the two distributions. But this is not sufficient in our
case. Recall that we are doing an exhaustive search on $2^{m_1}$
possible states, so the average number of wrong states passing the
statistical test will be $2^{m_1}2^\frac{-S}{2^{2n+1}}$. We thus need
a number of samples equal to at least $m_1 2^{2n+1}$ to obtain only a
few possible candidates for the real initial states. Notice that with
this number of samples, the probability to miss the initial states is
of $O(2^{-m_1})$ which is really small.

What is the amount of keystream needed to get that many samples? For
one sample, we will need to consider $2^{n_1}$ degree 4 equations
since this is the expected number for
$\yy_t+\yy_{t+t_1}+\yy_{t+t_2}+\yy_{t+t_3}$ to be equal to $\zero$.
The average degree of the lowest weight 4 multiples is
$O(2^\frac{m_2}{3})$ (see previous subsection).
By shifting this multiple $x$ times, we get $x$ degree 4 equation for
$O(2^\frac{m_2}{3} + x)$ consecutive keystream bits.
Hence, the required amount of keystream is $O(2^\frac{m_2}{3} +
m_1 2^{2n+n_1+1})$.

We assumed here that we use only one weight 4 multiples. If we use
many, it is actually possible to need less keystream at the expense
of more precomputation. This gain is difficult to analyze as it
really depend on the used LFSRs, but may definitely be useful in
practice. However, it will not change the overall asymptotic
keystream needed.

\subsection{Performing the attack efficiently}

We show in this subsection how to find the initial states in $O(m_1
2^{n_1} 2^{m_1})$ instead of $O(N 2^{m_1})$ with a straightforward
implementation. Notice that this is a huge gain since in our case
$m_1 2^{n_1} \ll N$. On the memory side, we need $2^{m_1}$ integer in
the second case compared to $N$ bits in the first one. If this is a
problem, we can actually trade memory for time and be anywhere
between those two algorithms.

In order to perform our attack, we have to compute many quantities
involving time positions of the form $(t,t+t_2,t+t_2,t+t_3)$. Since
we may use many multiples, let just assign an index $i$ to such
$4$-tuple. We will then write $z(i)$ for the sum of the
$z_t$ for the 4 time positions number $i$ and $\yy(i)$ in the same
way.

Computing the probability estimate for a given partial initial state is roughly the same as computing the number of indices $i$ such that $\yy(i)=\zero$ and $z(i)=0$.
Actually to get the true probability, we also need to know how many indices are such that $\yy(i)=\zero$ and $z(i)=1$ but this will not change our discussion or the final complexity.
So let restrict ourselves on the $N'$ indices $i$ such that $z(i)=0$.

If we do this independently for each of the $2^{m_1}$ partial initial
states, the complexity is then in $N' 2^{m_1}$ which is pretty large. In
order to improve on this complexity, let us start by assuming that out
of the $m_1$ bits, only one is used as an input for the filtering
function $f$, that is we assume the $\yy_t$ to be scalar. Remark now
that any linear combination of bits from the internal states of some LFSRs can
be expressed as a linear expression of the initial state of theses
LFSRs. This is the case for the $\yy(i)$. We can then define a binary
linear code of generator matrix $G$ of size $m_1 \times N'$ such that
for a given partial initial state $\uu$ the $i$-th element of $\uu G$
is precisely $\yy(i)$. The number we try to compute for a given
initial state $\uu$ is then just $N'$ minus the Hamming weight of
$\uu G$.

Is there a way to compute the Hamming weight of each codeword in a
code of length $N'$ and dimension $m_1$ faster than $2^{m_1} N'$ ? The
answer is yes, thanks to the Walsh transform we can do it in $O(m_1
2^{m_1})$ which in our case is a lot better since $m_1 \ll N'$. The
Walsh transform $\widehat{w}$ of a function $w: \F_2^{m_1} \to \Z $
can be computed in $O(m_1 2^{m_1})$ and is such that
$$ \widehat{w} (\uu) = \sum_{\vv} w(\vv)(-1)^{\vv.\uu} \ .$$
If $w(\vv)$ is equal to the number of columns in $G$ equal to $\vv$,
$\widehat{w}(\uu)$ is exactly twice the Hamming weight of $\uu G$
minus $N$.

Now, what to do when the $\yy_t$ are not scalar ?
we can use this nice formula :
$$
\#\{i, \yy(i)=\zero \} = \frac{\sum_{\yy\in\F_2^{n_1}} \#\{i, \yy(i).\yy=0\} - 2^{n_1-1} }{2^{n_1}} \ .
$$
This simply comes from the fact that when $\yy(i)$ is $\zero$ it
contributes to $2^{n_1}$ in the sum whereas any other $\yy(i)$
contributes only $2^{n_1-1}$. Each cardinality in the sum can be
computed using a Walsh transform, but since Walsh transform is
linear, we better compute directly the Walsh transform of the
function
$$
    w(\vv) = \#\{ (i,\yy), \   \yy.(G_1(i),\dots,G_{n_1}(i)) = \vv \}
$$
where $G_j(i)$ is the $i$-th columns of the matrix of the linear code corresponding to the input bit $j$ of the filtering function.

Finally, the time-memory tradeoff mentioned in the first paragraph
directly follows from a time memory tradeoff in the implementation of
the Walsh transform.

\section{Attack Example}\label{example}

We give in this section an example of how to use our result to mount an attack
on a given combination generator.

The following timings were obtained on an Intel Core2 Quad CPU Q9550
at 2.83GHz, using only one core and no more than 2GB of memory. We
used a combination  generator based on three  LFSRs  of size  $29$,
$31$  and  $37$ respectively.  The  feedback  polynomials  are  dense
in  order  not  to  have artificially easy to find low weight
multiples. The filtering function $f$ is a $9$-variable  Boolean
function,  with $3$ inputs from each LFSR. It was chosen to be still
balanced even if we fixed the input bits from any of the constituent
LFSR in order to avoid traditional correlation attack
, i.e. the function is $3$-resilient.

\begin{table}[ht]
  \centering
  \begin{tabular}{|c||c|c|c|}
  \hline
  &Precomputation & Total online time & Keystream used\\
  \hline
  \hline
  Attack 1 & 12min27s & 7min01s& 3.06MB \\
  \hline
  Attack 2 & 3min02s & 6h18min & 985KB \\
  \hline
  \end{tabular}
  \caption{Global comparison of the two attacks}
\label{table1}
\end{table}

\begin{table}[ht]
  \centering
  \begin{tabular}{|c||c|c|c|}
  \hline
  &1st LFSR & 2nd LFSR & 3rd LFSR\\
  \hline
  \hline
  Attack 1 & 51s & 6min10s & 0s\\
  \hline
  Attack 2 & 6h17min & 1min27s & 0s\\
  \hline
  \end{tabular}
  \caption{Comparison of the different parts of the online attacks}
\label{table2}
\end{table}

In the attack referred below as  Attack 1, we retrieve the initialization of the
LFSR of size  $29$, $31$ and $37$ in  this order. We thus have in  a first step
the following parameters~:
\[m_1=29;\; m_2=31+37=68;\; n_1=3;\; n_2=6.\]
As stated in Lemma \ref{lemma1}, we thus need to consider approximately $N=2^{26.86}$
  multiples of $P_{31}\times P_{37}$ of weight 4.
The maximum degree needed is slightly less than $2^{25}$ which implies we need
3MB of keystream.
Using the approach of \cite{DLC07} for finding low weight multiples
of degree less than $2^{25}$, the precomputation took around $13$ minutes. The online
time to recover the initial internal state of the first LFSR is only
$51s$, for a theoretical workload of approximately $2^{37}$.
The internal state of the second  LFSR is then recovered in roughly $6$ minutes,
using only very few keystream because we only need to consider multiples of the
last feedback polynomial.
Finally, the internal state of the  last register is found by a different method
almost  instantly, because  the  full knowledge  of  6 entries  of our  Boolean
function amongst 9 gives us a  lot of information.

In the second version of the attack , we retrieve the initialization of the
LFSR of size  $37$, $29$ and $31$ in  this order. We thus have in  a first step
the following parameters~:
\[m_1=37;\; m_2=31+29=60;\; n_1=3;\; n_2=6.\]
This small difference allows us to  have a lower value for $m_2$, which implies
that  we  need less  keystream  to  perform the  attack.  We  need to  consider
approximately $N=2^{26.86}$ multiples of $P_{31}\times P_{29}$ of weight 4,
the precomputation is this time only $3$ minutes and the maximum degree
needed is approximately $2^{23}$, corresponding to 985KB.
On the contrary,  the online time to recover the initial  internal state of the
first  LFSR   is  much  longer.  The  theoretical   workload  is  approximately
$2^{42}$ and  our experiment  confirms this ratio  as it tooks us just more  than 6
hours.
The timings to recover the two  other LFSR is negligible compared to the first
one.

In conclusion, the  Attack 1 is the  best one from a complexity  point of view,
whereas the Attack 2 minimizes the amount of keystream needed.
We  summarize the timings  for the  two different  strategies in Tables
\ref{table1} and \ref{table2}.


\section{Conclusion}\label{conclusion}

We presented in this paper an efficient attack on the combination
generator. In particular, if we look at the timings given in Section
\ref{example}, we are not aware of any other attacks that can break
the chosen combination generator that efficiently. Remark however
that breaking the combination generator still requires an exponential
number of computation. Hence, if the parameters are chosen large
enough, such a system can still be secure given the actual knowledge.

An important point is that the presented attack is inherent to the
construction and  is not based on any particular weakness in the
choice of the filtering function or in the constituent LFSRs. As
such, it appears that for the same order of internal size the
combination generator is a lot less secure than a single large
primitive LFSR filtered by a non-linear function. This statement
seems true in many ways. With a large LFSR, it is more difficult to
compute  any low weight multiples, it is more difficult to break the
system into smaller components, and the best attacks we know are a lot
less efficient.

%

\section*{Acknowledgment}

We would like to thank Anne Canteaut and Jean-Pierre Tillich for their help and
their useful comments.

\bibliography{article}
\bibliographystyle{alpha}

\appendix
\section{Proof of Theorem \ref{thm}}

The bound on $P_0$, mentioned in \cite{Canteaut06}, is already present in \cite{Lev04}.
It is a direct consequence of the result in \cite{Did07} where it is shown that
\begin{equation*}
P_\xx =
\frac{1}{2}\left(
1 + \sum_{\yy \in \F_2^n} (-1)^{\yy.\xx} \left(\frac{W_f(\yy)}{2^n}\right)^4
\right)
\end{equation*}
where $W_f(\yy)$ is the Walsh coefficient of $f$ at point $\yy$, that is $\sum_{\xx} (-1)^{f(\xx)+\xx.\yy}$.
By Parceval equality we know that the sum $\sum_{\yy} W_f(\yy)^2$ is equal to $2^{2n}$ and it is well know that the sum of square $\sum_{\yy} \left(W_f(\yy)^2\right)^2$ is minimized when every terms are equal.
Hence we can upper bound everything by
\begin{equation*}
\frac{1}{2}\left(1+
\frac{1}{2^{4n}} \sum_{\yy\in\F_2^n} \left(2^n\right)^2
\right)
\end{equation*}
and get the first part of the theorem.

For the second part, the result is taken directly from \cite{Did07}.
Just recall that the maximum of the autocorrelation coefficient is defined as
\begin{equation*}
\Delta_f \eqdef \max_{\yy \ne \zero} \left|\sum_{\xx \in \F_2^n} (-1)^{f(\xx)+f(\xx+\yy)}\right|
\end{equation*}
and is usually quite small for Boolean functions used in cryptography.

\end{document}